\documentclass[preprint,prd,showpacs,amssymb,floatfix,12pt]{revtex4}

\usepackage{graphicx}
\usepackage[latin1]{inputenc}
\usepackage{amsmath}
\usepackage{amsfonts}
\usepackage{amssymb}
\usepackage{amsthm}
\usepackage{bm}

\begin{document}
\input{epsf}




\title{Brownian Motion in Robertson-Walker Space-Times from
Electromagnetic Vacuum Fluctuations }

\author{Carlos H. G. B\'essa}
\email{carlos@cosmos.phy.tufts.edu }
\affiliation{Institute of Cosmology, Department of Physics and
Astronomy, \\Tufts University, Medford, MA 02155 USA}

\affiliation{Departamento de F\'{\i}sica, 
Universidade Federal da Para\'{\i}ba, 
\\Jo\~ao Pessoa, PB 58051-970, Brazil}

\author{Valdir B. Bezerra}
\email{valdir@fisica.ufpb.br}
\affiliation{Departamento de F\'{\i}sica, 
Universidade Federal da Para\'{\i}ba, 
\\Jo\~ao Pessoa, PB 58051-970, Brazil}

\author{L. H. Ford}
\email{ford@cosmos.phy.tufts.edu}
\affiliation{Institute of Cosmology, Department of Physics and
Astronomy, \\Tufts University, Medford, MA 02155 USA}

\begin{abstract}

We consider classical particles coupled to the quantized
electromagnetic field in the background of a spatially flat
Robertson-Walker universe. We find that these particles typically
undergo Brownian motion and acquire a non-zero mean squared velocity
which depends upon the scale factor of the universe. This Brownian
motion can be interpreted as due to non-cancellation of
anti-correlated vacuum fluctuations in the time dependent background
space-time. We consider several types of coupling to the
electromagnetic field, including particles with net electric charge,
a magnetic dipole moment, and electric polarizability. We also
investigate several different model scale factors.

\end{abstract}


\maketitle

\baselineskip=12pt

\section{Introduction}

Brownian motion of a particle in a thermal bath is a well-known
phenomenon. (See, for example, Ref.~\cite{Pathria}.) 
In this case, the particle's mean squared velocity grows
linearly in time until dissipation effects become important, after
which it approaches a non-zero equilibrium value. The linear growth
phase is characteristic of any random walk process, in which each
fluctuation is independent of previous fluctuations. Quantum
fluctuations are quite different from thermal ones in that the former
are strongly correlated or anti-correlated. This does not, however,
prevent quantum Brownian motion, which will be the topic of this
paper. The existence of Brownian motion in the Minkowski vacuum state 
is controversial. Although conventional quantum electrodynamics
suggests that the only effect will be an unobservable mass 
renormalization, Gour and Sriramkumar~\cite{gs99} have argued that there
could be an observable effect on charged particles coupled to the fluctuating
electromagnetic field. Brownian motion in the presence of boundaries
is less controversial, and has been studied by several 
authors~\cite{b9191,jr92,wkf02,yf04,wl05,sw07}. 
Barton~\cite{b9191} was the first to examine
fluctuations of the Casimir force. Wu {\it et al}~\cite{wkf02}
calculated the Brownian motion of an atom near a perfectly reflecting
plate due to fluctuations in the retarded van der Waals force. The
mean force here is the Casimir-Polder force~\cite{cp48}.
The analogous Brownian motion of a charged particle near a reflecting
plate was treated by Yu and Ford~\cite{yf04}. In all of these
cases, the mean squared velocity of the particle approaches a
constant even in the absence of dissipation. This is required by
energy conservation, as there is no energy source in these static 
configurations. The fact that the late time  mean squared velocity
is non-zero can be attributed to the effects of switching when
the interaction is turned on. Switching effects were recently discussed
by Seriu and Wu~\cite{sw07}.   

The mechanism which enforces the lack of growth of the  mean squared
velocity can be understood as anti-correlated fluctuations. A
charged or polarizable particle in a Casimir vacuum can acquire an
energy $E$ from a fluctuation. However, that energy is typically
surrendered on a time scale of order $\hbar/E$ to an
anti-correlated fluctuation. The correlation functions of the
quantized electromagnetic field automatically enforces the required
anti-correlations~\cite{ford07}. The quantum fluctuations of the
stress tensor in flat spacetime also exhibits subtle correlations and 
anti-correlations, as is discussed in Ref.~\cite{fr05}. 

The Brownian motion of test particles is an operational means to
describe a fluctuating quantum field. This approach can be used 
to treat the quantum fluctuations of the gravitational field,
which has been a topic of much interest in recent 
years~\cite{f82,kf93,f05,fs96,fs97,nd00,c98,ch97,ccv97,nbm1998,mv99,
hs98,tf06,wnf07}.

In the present paper, we will investigate the Brownian motion of
various types of particles coupled to the quantized electromagnetic
field in the background of a Robertson-Walker spacetime. Here
the time-dependent background geometry can act as an energy source,
so the particles can acquire a net kinetic energy. 

In Sect.~\ref{sec:formalism}
we develop the basic Langevin equation formalism for calculating 
the mean squared velocity of classical particles coupled to a 
fluctuating force in  a spatially flat Robertson-Walker background.
The formalism is applied to a several specific choices for the scale
factor of the universe in Sect.~\ref{sec:specific}. 
Our results are summarized and discussed in Sect.~\ref{sec:final}.

Unless otherwise noted, we work in Lorentz-Heaviside units with
$\hbar = c =1$.

\section{Basic Formalism}
\label{sec:formalism}

The equation of motion of a classical point particle moving in a curved 
spacetime with a four-force $f^{\mu}$ is

\begin{equation}\label{eqforce1}
f^{\mu} = m\frac{Du^{\mu}}{d\tau}\, ,
\end{equation}
where $u^{\mu}$ is the 4-velocity of the particle, $m$ is its mass and
$\tau$ the proper time. 
The operator $D/d\tau$ on the right-hand side of Eq. (\ref{eqforce1})
is the covariant derivative given by

\begin{equation}
\frac{Du^{\mu}}{d\tau} = \frac{du^{\mu}}{d\tau} +
\Gamma^{\mu}_{\alpha\beta}u^{\alpha}u^{\beta} \,.
\end{equation}

Here we take the space-time geometry to be that of a spatially flat
Robertson-Walker universe with metric
\begin{equation}\label{eqrwcom1}
ds^2 = -dt^2 + a^2(t)(dx^2 + dy^2 + dz^2) \,,
\end{equation}
where $a(t)$ is the scale factor. We will restrict our attention to
the case where the particles 
are moving slowly with respect to these coordinates, in which case the
particle's proper time 
becomes the coordinate time $t$. Because of spatial isotropy, we can
consider a particular direction, the $x$-direction, and write the
equation of motion as 
\begin{equation}\label{eqbasic}
\frac{du^{x}}{dt} + 2\frac{\dot{a}}{a}u^x = \frac{1}{m}f^{x}\,.
\end{equation}
Here we have used $\Gamma^{x}_{tx} =\Gamma^{x}_{xt} = \dot{a}/a$,
 where  $\dot{a} = da/dt$.
We will take the four-force to be of the form
\begin{equation}
f^x = f^{'x} + f^x_{ext}\,,
\end{equation}
where $f^x_{ext}$ is a non-fluctuating external force, and $f'^x$ is a
fluctuating force produced 
by the electromagnetic vacuum fluctuations whose mean value vanishes:
\begin{equation}
\langle f^{'x} \rangle =0 \,.
\end{equation}

First, let us consider the case of free particles, which corresponds
to the case when
 $f_{ext} = 0$. Thus Eq. (\ref{eqbasic}) can be written as,
\begin{equation}
\frac{1}{a^2}\frac{d}{dt}\left(a^2u^x\right) = \frac{f^{'x}}{m},
\end{equation}
which after integration reduces to
\begin{equation}
a^2(t_f)u^x(t_f) - a^2(t_0)u^x(t_0) = \frac{1}{m}\int_{t_0}^{t_f} dt
a^2(t)f^{'x}(t).
\end{equation}
Assuming that these particles are initially  at rest ($u^x(t_0) = 0$),
we find that the 
velocity-velocity correlation function is given by
\begin{equation}\label{eqbasicf}
\langle u^x(t_f, r_1)u^x(t_f, r_2)\rangle = \frac{1}{m^2a_f^4}\int
dt_1dt_2a^2(t_1)a^2(t_2)\;\langle f^{'x}(t_1, r_1)f^{'x}(t_2, r_2)\rangle \,.
\end{equation}

Another case of interest is when there is an external force which
cancels the effect of the cosmological expansion:
\begin{equation}
 f^x_{ext} =  2\frac{\dot{a}}{a}\, u^x \,.
\end{equation}
This is the case for any particles in bound systems such as galaxies
or molecules. 
Such particles do not participate in the cosmological expansion and in
this case two such 
particles do not move apart on the average.
We will refer to these as bound particles. In this case, 
\begin{equation}
\frac{du^{x}}{dt}  = \frac{1}{m}f'^{x}\,,
\end{equation}
and the velocity correlation functions for particles which start 
at rest at $t=t_f$ is

\begin{equation}\label{eqbasicb}
\langle u^x(t_f, r_1)u^x(t_f, r_2)\rangle = \frac{1}{m^2}\int
dt_1dt_2 \, \langle f^{'x}(t_1, r_1)f^{'x}(t_2, r_2)\rangle \,.
\end{equation}

Note that the above expression is a coordinate velocity correlation
function. In a Robertson-Walker space-time, proper distance between
particles,  $l_f$, is 
related to the coordinate separation $r$ at $t=t_f$ by $l_f =
a_fr$. Thus the proper velocity 
correlation function is given by

\begin{equation}
\langle v^x(t_f,r_1)v^x(t_f, r_2)\rangle = 
a_f^2 \langle u^x(t_f, r_1)u^x(t_f,r_2)\rangle \,.
\end{equation}

\subsection{Charged Particles}

In this section, we will consider electrically charged particles with
charge $q$ coupled to a 
fluctuating electromagnetic field. In this case, the four-force is
\begin{equation}
f'^x = \frac{q}{m}F^{xt}u_t \approx - \frac{q}{m}F^{xt} \, .
\end{equation} 
For the case of free particles,  Eq. (\ref{eqbasicf}) yields
\begin{equation}\label{eq1ele}
\langle u^x(t_f, r_1)u^{x}(t_f, r_2)\rangle =
\frac{q^2}{4a_f^4}\int dt_1 \int dt_2\; a^2(t_1)a^2(t_2)\,
\langle\left(F^{xt}u_t\right)_1\left(F^{xt}u_{t}\right)_2\rangle_{RW},
\end{equation}
where the sub-indexes 1 and 2 refer to the coordinates $(t_1, r_1)$
and $(t_2, r_2)$, 
respectively, and the subscript $RW$ denotes a vacuum correlation
function in 
Robertson-Walker spacetime.

This correlation function is obtained from the corresponding
correlation function in flat 
spacetime by a conformal transformation. First write the
Robertson-Walker metric in its conformal form
\begin{equation}
ds^2 = a^2(-d\eta^2 + dx^2 + dy^2 + dz^2),
\end{equation}
with $dt = ad\eta$. The field strength tensor in these coordinates is given by
\begin{equation}
\left(F^{\mu\nu}\right)_{RW} =a^{-4}\left(F^{\mu\nu}\right)_{M}\, ,
\label{eq:conformal}
\end{equation}
where the subscript $M$ refers to the Minkowski space field
strength. This may be seen, for example, from the fact that the
Lagrangian density, $\sqrt{-g} F^{\mu\nu}F_{\mu\nu}$ is invariant
under the conformal transformation. 
From this and Eq. (\ref{eq1ele}), we find

\begin{equation}\label{eq3ele}
\langle u^x(t_f, r_1)u^{x}(t_f, r_2)\rangle =
\frac{q^2}{m^2a_f^4}\int d\eta_1 \int d\eta_2 \,
\langle F^{\eta x}(\eta_1, r_1)F^{\eta x
}(\eta_2,r_2)\rangle_{M}\,.
\end{equation}
Here the appropriate component of the Minkowski space correlation
function is given 
by Eq.~(\ref{eq:ExEx}).
The key feature of this result is that the scale factor does not
appear inside the integrand. 
Thus the cosmological expansion has no effect on the Brownian motion, and
hence we do not find an interesting result in this case.

The case of bound charged particles is different. In this case, from 
Eqs.~(\ref{eqbasicb}) and (\ref{eq:ExEx}), we find
\begin{eqnarray}\label{eq3}
\langle  u^x(\eta, r_1)u^x(\eta, r_2) \rangle =
\frac{q^2}{m^2}\int d\eta_1\int d\eta_2  \langle
\left(F^{tx}u_t\right)_1 \left(F^{tx}u_t\right)_2\rangle_{RW}
\\ \nonumber
= \frac{q^2}{m^2}\int d\eta_1\int d\eta_2\;
a^{-2}(\eta_1)a^{-2}(\eta_2) \,
\left\{\frac{-(\eta_2 - \eta_1)^2 - r'^2}{\pi^2[-(\eta_2 -
\eta_1)^2 + r^2 ]^3}\right\}\, .
\end{eqnarray}
Here $r$ is the spatial separation of the particles and $r'^2 =r^2
-2\Delta x^2$. Now there are factors of $1/a^2$ in the integrand, 
which will lead
to non-trivial effects in an expanding universe.

\subsection{Magnetic Dipoles}

In this section we will consider particles with a magnetic dipole
moment. In flat spacetime, such particles experience a force when
there is a non-zero magnetic field gradient:
\begin{equation}
\vec{f} = -\nabla u
\end{equation}
where $u = - \vec{\mu}.\vec{B}$, is the magnetic potential energy, $\mu$ is the magnetic moment and $\vec{B}$ is the magnetic field. Writing this force in covariant form, we have for the $x$-component

\begin{equation}
f^{x} = \mu \partial^x \left(F^{zy}x_zx'_y\right)
\end{equation}
where, $x_{\mu} = (0, 0, 0, a)$ and $x'_{\mu} = (0, 0, a, 0)$, and
such that $x^{\mu}x_{\mu} = x'^{\mu}x'_{\mu} = 1$.

For free particles, the velocity-velocity correlation function, 
Eq.~(\ref{eqbasicf}), becomes
\begin{equation}
\langle u^{x}(t_f, r_1)u^{x}(t_f, r_2)\rangle =
\frac{\mu^2}{a^4(t_f)m^2}\int dt_1\int dt_2
\partial_{x_1}\partial_{x_2} \,
\langle\left(F^{zy}x_{z}x'_{y}\right)_{1}
\left(F^{zy}x_{z}x'_{y}\right)_{2}\rangle_{RW}.
\end{equation}
Again, we may use Eq.~(\ref{eq:conformal}) to write the above
expression in terms of 
$\partial_{x_1}\partial_{x_2}\langle F^{zy}(\eta_1, r_1)\,
F^{zy}(\eta_2, r_2) \rangle_{M}$, 
which may be evaluated to write
\begin{eqnarray}\label{eqmag2}
\langle u^{x}(\eta,r_1)u^x(\eta,r_2)\rangle  =
-\frac{\mu^2}{m^2a^4(t_f)}\int d\eta_1\int d\eta_2
a^{-1}(\eta_1)a^{-1}(\eta_2) \times \\ \nonumber
\left\{\frac{2}{\left[-(\eta_2 - \eta_1)^2 + r^2\right]^3} +
\frac{6(\eta_2 - \eta_1)^2}{\left[-(\eta_2 - \eta_1)^2 +
r^2\right]^4} \right\}.
\end{eqnarray}
In Eq. (\ref{eqmag2}), we have used the  coincident limit $r
\rightarrow 0$ only 
in the numerator, in order to simplify our expressions. This will not
alter our final results, because we will take this  limit 
after the integrations.

For bound magnetic dipoles, we may start with Eq.~(\ref{eqbasicb}) and
follow the same procedure to find
\begin{eqnarray}\label{eq2mag2}
\langle u^x(\eta,r_1)u^x(\eta,r_2)\rangle && =
-\frac{\mu^2}{m^2}\int
d\eta_1\int d\eta_2 a^{-3}(\eta_1)a^{-3}(\eta_2) \times \\
\nonumber && \left\{\frac{2}{\left[-(\eta_2 - \eta_1)^2 +
r^2\right]^3} + \frac{6(\eta_2 - \eta_1)^2}{\left[-(\eta_2 -
\eta_1)^2 + r^2\right]^4} \right\},
\end{eqnarray}
where again the coincidence limit in the spatial coordinate was taken
in the numerator of the integrand.

\subsection{Polarizable Particle}

We will consider in this section a polarizable particle, described as
a point particle with a static polarizability $\alpha$. In an 
inhomogeneous electric field, such a
particle experiences a force
\begin{equation}
\vec{f}(x) = \frac{\alpha}{2} \nabla(E^2) \,,
\end{equation}
which in covariant notation becomes,
\begin{equation}\label{eqreview1}
f^x=
\frac{\alpha}{2}\partial^x\left(F^{tx}u_tx_x\right)^2,
\end{equation}
where the low velocity limit was taken with 
$u_{\nu} = -\delta^0_{\nu}$, $x_{\mu} = (0, a, 0, 0)$. For a free
particle, Eqs.~(\ref{eqbasicf}) and (\ref{eqreview1}) lead to
\begin{eqnarray}\label{eq3pol}
\langle u^x(t_f, r_1)u^{x}(t_f, r_2)\rangle &&=
\frac{\alpha^2}{4a_f^4}\int d\eta_1 \int d\eta_2 \times  \\
\nonumber &&a^{-3}(\eta_1)a^{-3}(\eta_2)
\partial_{x_1}\partial_{x_2}\langle\left[F^{0i}(\eta_1,
r_1)\right]^2\left[F^{0i}(\eta_2,r_2)\right]^2\rangle_{M},
\end{eqnarray}
where we used the fact that $u_{\mu} = -a\delta_{\mu}^{\eta}$ and
$x_{\mu} = (0, a, 0, 0)$, in conformal coordinates.  

Here we use the Wick theorem to calculate the two-point function:
 $\langle E^2(r_1)E^2(r_2)\rangle_M = 
\langle E_i(r_1)E^i(r_2)E_j(r_1)E^j(r_2)\rangle_M$, finding 

\begin{equation}
\langle E_i(r_1)E^i(r_2)E_j(r_1)E^j(r_2)\rangle_M = 2\left[\langle
E_i(r_1)E_j(r_2)\rangle_M \langle
E^i(r_1)E^j(r_2)\rangle_M\right],
\end{equation}
or, from the procedure outlined in Appendix A,
\begin{equation}
\langle E_i(r_1)E^i(r_2)E_j(r_1)E^j(r_2)\rangle_M =
\left\{\frac{-3(\eta_2 - \eta_1)^2 - r^{2}}{\pi^2[-(\eta_2 -
\eta_1)^2 + r^2 ]^3}\right\}^2 \,.
\end{equation}
To simplify our expression we will consider the coincident limit in
the spatial
coordinate $r = r_2 - r_1 = 0$, only in the numerator of all factors,
as in the previous cases. Then the velocity-velocity correlation function is,
\begin{eqnarray}
&&\langle u^x(\eta, r_1)u^x(\eta, r_2)\rangle =
\frac{\alpha^2}{4m^2a_f^4}\int
d\eta_1 \int d\eta_2 a^{-3}(\eta_1)a^{-3}(\eta_2) \times \\
\nonumber &&\left\{ \frac{32}{[-(\eta_2 - \eta_1)^2 + r^2]^5} +
\frac{136(\eta_2-\eta_1)^2}{[-(\eta_2 - \eta_1)^2 + r^2]^6} +
\frac{144(\eta_2 - \eta_1)^4}{[-(\eta_2 - \eta_1)^2 + r^2]^7}
\right\}.  \label{eq:pol}
\end{eqnarray}

\section{Specific Universe Models}

\label{sec:specific}

In this section we will apply the basic formulas obtained in 
Sect.~\ref{sec:formalism} to
investigate the influence of different scale factors on the Brownian
motion of particles induced by quantum vacuum fluctuations 
of the electromagnetic field.

\subsection{Asymptotically Static Bouncing Universe}
\label{sec:bounce1}

The study of bouncing universe was considered by some authors in the
past~\cite{as78}. 
Here we will study a special case, which is asymptotically 
static in the past and future.
We will take the scale factor to have the form
\begin{equation}\label{eq77}
a^n=\frac{\eta^2+\eta_0^2}{\eta^2+G^2\eta_0^2},
\end{equation}
where $G$ and $\eta_0$ are constants, and $n$ is a positive
integer. Note that when
$\eta \rightarrow \pm\infty$, the universe is asymptotically flat and
$a$ goes to unity. 
It will be convenient to consider different choices of $n$ for
different types of particles 
in order to simplify the corresponding integrals.

\subsubsection{Bound Charged Particles}

In this case, we set $n=1$, so that Eq.~(\ref{eq3}) becomes
\begin{eqnarray}\label{eq5}
\langle \Delta u(\eta, r_1) \Delta u(\eta, r_2)\rangle =
\frac{q^2}{m^2} \int_{-\infty}^{+\infty}d\eta_1
\int_{-\infty}^{+\infty}d\eta_2
\left(\frac{\eta_1^2+G^2\eta_0^2}{\eta_1^2+\eta_0^2}\right)^2
\\ \nonumber  \times \left(\frac{\eta_2^2+G^2\eta_0^2}{\eta_2^2+
\eta_0^2}\right)^2  \left[\frac{-(\eta_2-\eta_1)^2 -
r^{'2}}{\pi^2[-(\eta_2-\eta_1)^2 +  r^2]^3}\right],
\end{eqnarray}
where we used Eq. (\ref{eq3}). This integral is evaluated in Appendix
B, with the result 
being  a rather complicated expression, Eq.~(\ref{eq:delu1}).
This result simplifies considerably in the limit that 
$r_1\rightarrow r_2$ and $G \gg 1$ to
\begin{equation}\label{eq8}
\langle\Delta u^2 \rangle = \frac{21q^2G^8}{128m^2\eta_0^2}.
\end{equation}
Note that because $a_f=1$, this expression also gives the mean squared
proper velocity $\langle\Delta v^2 \rangle$.

We can gain some insight into this result by writing it in terms of a
characteristic measure of the maximum curvature.  Consider the scalar
curvature $R$  and evaluate it using the scale factor Eq.~(\ref{eq77})
with $n=1$. 
We find the following result,
\begin{equation}\label{eqr010}
R_0 = \frac{12G^2}{\eta_0^2}\left(1-G^2\right),
\end{equation}
where $R_0$ is the Ricci scalar when $\eta = 0$. If we consider the limit $G
\gg 1$, $R_0$ is negative and
\begin{equation}\label{eq9}
R_0^2 \approx \frac{144G^8}{\eta_0^4}.
\end{equation}
Then $\langle\Delta v^2\rangle$ in terms of $R_0$ is
\begin{equation}\label{100}
\langle\Delta v^2\rangle = \frac{7q^2\eta_0^2R_0^2}{6144m^2}.
\end{equation}

We can also write Eq.~(\ref{100}) in terms of the redshift, defined
by: $a_m^{-2} \equiv (1 + z)^2$, with  $a_m \equiv a(0)/a(\infty)
=(1/G)^2$, where $a_m$ is the minimum scale factor. 
Considering $z \gg 1$, we get 
\begin{equation}
\langle\Delta v^2\rangle = \frac{7q^2|R_0|}{6144m^2}z^2.
\label{eq:bcp1}
\end{equation}
The mean squared velocity is proportional both to the 
squared redshift and to the maximum curvature.
This  can be associated with an effective temperature using the
non-relativistic equation: 
$k_BT_{eff} = m \langle \Delta v^2\rangle$. Where $T_{eff}$ is the
effective temperature and 
$k_B$ is the Boltzmann constant in Lorentz-Heaviside units $\hbar
= c = 1$. Then, $T_{eff}$ is,

\begin{equation}\label{tb1}
T_{eff} \simeq
\frac{10^{-3}q^2}{k_B\lambda_c}\left(\frac{\lambda_c}{l_c}\right)^2z^2,
\end{equation}
where $l_c = 1/\sqrt{|R_0|}$ is the lenght curvature and $\lambda_c
=1/m$ is the particle's Compton wavelength.

\subsubsection{Free Magnetic Dipoles}

Again take the scale factor to be Eq.~(\ref{eq77}) with $n=1$. 
Then Eq.~(\ref{eqmag2}) for the mean squared velocity becomes,
\begin{eqnarray}
\langle \Delta u(r_1,\eta)\Delta u(r_2,\eta) \rangle &=&
\frac{\mu^2}{a_f^4m^2} \;
\int^{\infty}_{-\infty}d\eta_1\int^{\infty}_{-\infty}d\eta_2
\left(\frac{\eta^2 + G^2\eta_0^2}{\eta_1^2 + \eta_0^2}\right)
\left(\frac{\eta^2 + G^2\eta_0^2}{\eta_2^2 + \eta_0^2}\right)
\\ \nonumber   &\times&
\left\{\frac{-2}{[-(\eta_2 - \eta_1)^2 + r^2]^3} - 
\frac{6(\eta_2 -
\eta_1)^2}{[-(\eta_2 -\eta_1)^2 + r^2}\right\}.
\end{eqnarray}
Using the same procedure as before, we find the following result in
the coincident limit when $G \gg 1$
\begin{equation}
\langle \Delta v^2\rangle = \frac{
10^{-3}\mu^2}{m^2}\frac{|R_0|}{\eta_0^2}.
\end{equation}
In terms of the redshift ($z \gg 1$), we have 
\begin{equation}
\langle \Delta v^2\rangle = \frac{
10^{-3}\mu^2}{m^2}\frac{|R_0|^2}{z^2}.  \label{eq:fmd1}
\end{equation}
In contrast to the result for bound charges, the effect decreases with $z$. 
For the case of electrons, it is convenient to write the magnetic moment as
\begin{equation}
\mu \simeq \frac{q}{2 m}\, .
\end{equation}
The effective temperature in terms of curvature length and 
Compton wavelength $\lambda_c$  is,
\begin{equation}\label{tempmag1}
T_{eff} = \frac{
10^{-3}q^2}{\lambda_ck_B}\left(\frac{\lambda_c}{l_c}\right)^4z^{-2}.
\end{equation}

\subsubsection{Bound Magnetic Dipoles}

Here we choose the scale factor to be of the form of 
Eq.~(\ref{eq77}) with $n=3$.
In this case, the mean squared velocity  from Eq.~(\ref{eq2mag2}) is,
\begin{eqnarray}
\langle \Delta u(r_1,\eta)\Delta u(r_2,\eta) \rangle &=&
\frac{\mu^2}{m^2} 
\int^{\infty}_{-\infty}d\eta_1\int^{\infty}_{-\infty}d\eta_2
\left(\frac{\eta^2 + \eta_0^2}{\eta^2 + G^2\eta_0^2}\right)
\left(\frac{\eta^2 + \eta_0^2}{\eta^2 + G^2\eta_0^2}\right) 
\\ \nonumber &\times& \left\{\frac{-2}{[-(\eta_2 - \eta_1)^2 + r^2]^3} -
\frac{6(\eta_2 -
\eta_1)^2}{[-(\eta_2 -\eta_1)^2 + r^2}\right\}.
\end{eqnarray}
We may evaluate this integral using the same technique as before, with
the result in the coincident limit
\begin{equation}
\langle \Delta u^2 \rangle = \frac{6\times
10^{-2}\mu^2}{a_f^4m^2}\frac{\left(G^2 - 1\right)^2}{\eta_0^4}.
\end{equation}
The physical velocity when $G \gg H$ is,
\begin{equation}
\langle \Delta v^2\rangle = \frac{6 \times
10^{-2}\mu^2}{64m^2}|R_0|^3\eta_0^2,
\end{equation}
where  now the scalar curvature at $\eta = 0$ is
\begin{equation}\label{eq99}
R_0 = -\frac{4}{\eta_0^2}G^{4/3}.
\end{equation}
In terms of the redshift, given by 
$1+z  \approx z = a(\infty)/a(0) = G^{2/3}$, we can write
\begin{equation}
\langle \Delta v^2\rangle = \frac{ 10^{-2}\mu^2}{64m^2}|R_0|^2z^2,
\label{eq:bmd1}
\end{equation}
which shows that the effect of quantum fluctuations grows with $z$.
Associating an effective temperature we have,
\begin{equation}\label{eqtfbmp}
T_{eff} \simeq \frac{ 10^{-3}q^2}{
k_B\lambda_c}\left(\frac{\lambda_c}{l_c}\right)^4z^2.
\end{equation}
Here the temperature grows with $z$ as in the bounded electric
particle due the extra 
force that acts on the magnetic dipole. Indeed, we see that the effect
here is smaller than that one indicated by Eq. (\ref{tb1}).

\subsubsection{Free Polarizable Particle}

Again we take the scale factor to be of the form of Eq.~(\ref{eq77})
with $n=3$. Equation~(\ref{eq:pol}) for the mean squared
velocity can be written as
\begin{eqnarray}\label{eq55}
&&\langle \Delta u(r_1,\eta)\Delta u(r_2,\eta) \rangle =
\frac{\alpha^2}{4m^2a_f^4\pi^4} \times \\ \nonumber
&&\int_{-\infty}^{+\infty}d\eta_1\int_{-\infty}^{+\infty}d\eta_2
\left(\frac{\eta_1^2 + G^2\eta_0^2}{\eta_1^2 +
\eta_0^2}\right)\left(\frac{\eta_2^2 + G^2\eta_0^2}{\eta_2^2 +
\eta_0^2}\right)\times \\ \nonumber &&\left\{ \frac{32}{[-(\eta_2
- \eta_1)^2 + r^2]^5} + \frac{136(\eta_2-\eta_1)^2}{[-(\eta_2 -
\eta_1)^2 + r^2]^6} + \frac{144(\eta_2 - \eta_1)^4}{[-(\eta_2 -
\eta_1)^2 + r^2]^7} \right\}.
\end{eqnarray}
Following the procedure previously used, we find, in the coincident limit,
\begin{equation}\label{eq88}
\langle\Delta v^2 \rangle = \frac{3\times
10^{-2}\alpha^2}{4\pi^2m^2}\frac{\left(G^2 -
1\right)^2}{\eta_0^8}.
\end{equation}
This physical (comoving) velocity can be expressed in  terms 
of $R_0$ given in Eq.~(\ref{eq99}) as
\begin{equation}\label{10}
\langle\Delta v^2\rangle =
\frac{10^{-3}\alpha^2}{64\pi^2m^2}\frac{|R_0|^3}{\eta_0^2}.
\end{equation}
We can also write Eq.~(\ref{10}) in terms of the redshift as
\begin{equation}
\langle\Delta v^2\rangle =
\frac{10^{-3}\alpha^2}{256m^2\pi^2}|R_0|^4z^{-2}.
\end{equation}

The mean squared velocity decreases with the redshift in contrast with
the bounded particle 
cases investigated in the previous section. This is due to the fact
that the atoms are free of external forces. This effect can be 
associated with an effective
temperature using the non-relativistic equation: $k_BT_{eff} = m
\langle \Delta v^2\rangle$. Thus, we obtain
\begin{equation}\label{eqtpol}
T_{eff} \simeq
\frac{10^{-6}\alpha^2}{k_B\lambda_c}
\left(\frac{\lambda_c}{l_c^4}\right)^2z^{-2}.
\end{equation}
This result shows that the temperature decreases with the redshift
because the particles are free of external forces.

\subsection{Asymptotically Bounded Expansion}

\label{sec:bounded-expand}

A universe with asymptotically bounded expansion was  studied in
Ref.~\cite{bd77}, 
where the production of massive particles were considered. 
Here we will  investigate the
Brownian motion effects in scale factors of the form
\begin{equation}
a^n = a_0^n + a_1^n\tanh(\eta/\eta_0), \label{eq:abe}
\end{equation}
where $n$ is a positive integer, $a_0$ and $a_1$ are dimensionless
constants and $\eta_0$ is a constant with dimension of time. 
We note that when $\eta
\rightarrow \pm\infty \Rightarrow a^2 \rightarrow a_0^2 \pm a_1^2$. 
Then, this universe is asymptotically flat in past and future, but  it
is not symmetric and exhibits only expansion.

\subsubsection{Bound Charged Particles}

Here we take the scale factor to be given by $n=2$ in 
Eq.~(\ref{eq:abe}). The mean squared coordinate velocity is then given by

\begin{eqnarray}\label{eq667}
\langle\Delta u(\eta, r_1)\Delta u(\eta, r_2)\rangle =
\frac{q^2}{m^2} \int_{-\infty}^{+\infty} d\eta_2
\int_{-\infty}^{+\infty} d\eta_1 \left(\frac{1}{a_0^2 +
a_1^2\tanh\left(\frac{\eta_1}{\eta_0}\right)}\right)  \\
\nonumber \times\left(\frac{1} {a_0^2 +
a_1^2\tanh\left(\frac{\eta_2}{\eta_0}\right)}\right)
\left[\frac{-(\eta_2-\eta_1)^2 - r^{'2}}{\pi^2[-(\eta_2-\eta_1)^2
+ r^2]^3}\right].
\end{eqnarray}
This integral is calculated in Appendix B, with the result given 
by Eq.~(\ref{vel2}).

In this model, the physical velocity is related to the coordinate
velocity by $\langle\Delta v^2\rangle= a_f^2 \langle\Delta
u^2\rangle$, 
where $a_f^2 = a_0^2 +
a_1^2$. If we will make a Taylor expansion in $r$ up to the 
zeroth order term, we find
\begin{equation}\label{eq111}
\langle\Delta v^2\rangle = \frac{-4q^2(a_0^2 +
a_1^2)\sinh^4\left[\frac{1}{2}\ln\frac{\alpha^2 + 1}{\alpha^2 - 1
}\right]}{\pi^4 m^2 a_1^4 \eta_0^2}\left(9 - \frac{2\pi^4}{15}
+3\zeta(3) \right)\, ,
\end{equation}
where $\alpha = a_0/a_1$.
Here the expression in parenthesis is a negative constant and
$\zeta(x)$ is the 
Riemann zeta function. We can write Eq.~(\ref{eq111}) in terms of the  scalar
curvature at $\eta =0$, given by 

\begin{equation}
R_0 =\frac{6a_1^4}{\eta_0^2a_0^6}\,.
\end{equation}
It is also interesting write the mean squared velocity in terms of the
redshift  
defined here as $1+z \approx z = a(\infty)/a(-\infty$). 
Then, $\langle\Delta v^2\rangle$ in terms
of $R_0$ and the redshift is given by
\begin{equation}\label{eqsef}
\langle\Delta v^2\rangle \simeq 10^{-2}\frac{q^2}{m^2}R_0z^4,
\end{equation}
when $z \gg 1$.
The effective temperature is now
\begin{equation}\label{tse}
T_{eff} \simeq
10^{-2}\frac{q^2}{k_B\lambda_c}\left(\frac{\lambda_c}{l_c}\right)^2z^4.
\end{equation}

\subsubsection{Free Magnetic Dipole}

Now take the scale factor to be Eq.~(\ref{eq:abe}) with $n=1$. Then
the mean squared velocity is given by

\begin{eqnarray}
&&\langle \Delta u(r_1,\eta)\Delta u(r_2,\eta) \rangle =
\frac{\mu^2}{m^2a_f^4\pi^2} \times \\ \nonumber
&&\int_{-\infty}^{+\infty}d\eta_1\int_{-\infty}^{+\infty}d\eta_2
\left(\frac{1}{a_0 +
a_1\tanh(\eta_1/\eta_0)}\right)\left(\frac{1}{a_0 +
a_1\tanh(\eta_2/\eta_0)}\right) \\
\nonumber &&\times\left\{\frac{-2}{[-(\eta_2 - \eta_1)^2 + r^2]^3}
- \frac{6(\eta_2 - \eta_1)^2}{[-(\eta_2 -\eta_1)^2 + r^2}\right\},
\end{eqnarray}
Following the same procedure as before we find that in the 
coincidence limit $(r \rightarrow 0)$,
\begin{equation}
\langle \Delta u^2 \rangle = \frac{24\mu^2 R_0^2
}{\pi^6m^2a_f^4a_1^2\eta_0^4}\sinh^4\left[\frac{1}{2}\ln\left(\frac{\alpha
+ 1 }{\alpha - 1}\right)\right]\left[\zeta(5) - \zeta(6)\right].
\end{equation}
As before, $\alpha = {a_0}/{a_1}$.
The physical velocity is
\begin{equation}
\langle \Delta v^2 \rangle \simeq \frac{2 \mu^2 R_0^2 z^4
}{12\pi^6m^2},
\end{equation}
where now 
\begin{equation}
R_0 = \frac{6a_1^2}{a_0^4\eta_0^2}.
\end{equation}
The temperature in terms of $\lambda_c$ and $l_c$ is,
\begin{equation}
T_{eff} \simeq  10^{-3}\frac{q^2 }{k_B\lambda_c}
\left(\frac{\lambda_c}{l_c}\right)^{4}z^4.
\end{equation}

\subsubsection{Bound Magnetic Dipole}

In this case, let $n = 3$ in Eq.~(\ref{eq:abe}). The mean 
squared velocity becomes
\begin{eqnarray}
&&\langle \Delta u(r_1,\eta)\Delta u(r_2,\eta) \rangle =
\frac{\mu^2}{m^2\pi^2} \times \\ \nonumber
&&\int_{-\infty}^{+\infty}d\eta_1\int_{-\infty}^{+\infty}d\eta_2
\left(\frac{1}{a_0 +
a_1\tanh(\eta_1/\eta_0)}\right)\left(\frac{1}{a_0 +
a_1\tanh(\eta_2/\eta_0)}\right) \\
\nonumber &&\times \left\{\frac{-2}{[-(\eta_2 - \eta_1)^2 +
r^2]^3} - \frac{6(\eta_2 - \eta_1)^2}{[-(\eta_2 -\eta_1)^2 +
r^2}\right\}.
\end{eqnarray}
In the coincidence limit $(r \rightarrow 0)$,
\begin{equation}
\langle \Delta v^2 \rangle = \frac{3 \times 10^{-3} \mu^2 R_0^2
z^4 }{m^2}\left[\zeta(5) - \zeta(6)\right],
\end{equation}
where 

\begin{equation}
R_0 = \frac{2a_1^6}{a_0^8\eta_0^2}.
\end{equation}
The effective temperature is given by
\begin{equation}
T_{eff} \simeq 3 \times
10^{-3}\frac{q^2}{k_B\lambda_c}\left(\frac{\lambda_c}{l_c}\right)^4z^4.
\end{equation}

\subsubsection{Free Polarizable Particle}

The scale factor is again given by Eq.~(\ref{eq:abe}) with $n=3$ and
the mean squared coordinate velocity is given by
\begin{eqnarray}\label{eq66}
&&\langle\Delta u(\eta, r_1)\Delta u(\eta, r_2)\rangle =
\frac{\alpha^2}{4m^2a_f^4} \int_{-\infty}^{+\infty} d\eta_2
\int_{-\infty}^{+\infty} d\eta_1 \times  \\
\nonumber  &&\left(\frac{1}{a_0^3 +
a_1^3\tanh\left(\frac{\eta_1}{\eta_0}\right)}\right)
\left(\frac{1} {a_0^3 +
a_1^3\tanh\left(\frac{\eta_2}{\eta_0}\right)}\right) \times
\\
\nonumber &&\left\{ \frac{32}{[-(\eta_2 - \eta_1)^2 + r^2]^5} +
\frac{136(\eta_2-\eta_1)^2}{[-(\eta_2 - \eta_1)^2 + r^2]^6} +
\frac{144(\eta_2 - \eta_1)^4}{[-(\eta_2 - \eta_1)^2 + r^2]^7}
\right\} \,.
\end{eqnarray}
Following the method used previously, we find the result
\begin{eqnarray}\label{vel22}
\langle\Delta u^2\rangle =
\frac{10\alpha^4}{m^2a_f^4\pi^14\eta_0^8a_1^6}
\sinh^4\left[\frac{1}{2}\ln\left(\frac{\alpha^3+1}{\alpha^3-1}\right)\right]
\\ \nonumber \times \left[\zeta(9) - \zeta(10)\right],
\end{eqnarray}
in the limit $r \rightarrow 0$. Here $\zeta(x)$ is again the zeta
function, and the  expression in brackets is positive.
The physical speed satisfies the relation
\begin{eqnarray}\label{eq11}
\langle\Delta v^2\rangle =
\frac{10\alpha^4}{m^2a_f^2\pi^{14}\eta_0^8a_1^6}
\sinh^4\left[\frac{1}{2}\ln\left(\frac{\alpha^3+1}{\alpha^3-1}\right)\right].
\end{eqnarray}
We can write Eq.~(\ref{eq11}) in terms of the  scalar curvature $R_0 =
{2a_1^6}/(\eta_0^2a_0^8) $, and of the redshift  as
\begin{equation}\label{eqsef2}
\langle\Delta v^2\rangle \simeq
10^{-11}\frac{\alpha^2}{m^2}R_0^4z^4,
\end{equation}
when $z \gg 1$.
The corresponding effective temperature is
\begin{equation}\label{tempboun1}
T_{eff} \simeq
10^{-11}\frac{\alpha^2}{k_B\lambda_c}\left(\frac{\lambda_c}{l_c^4}\right)^2z^4.
\end{equation}
Note that for this class of scale factors, we find the unexpected
result that 
$\langle\Delta v^2\rangle \propto z^4$ for all four types of 
particles being considered.
These results will be discussed in more detail in Sect.~\ref{sec:final}.

\subsection{Another Bouncing Universe}

\label{sec:bounce2}

Here we will consider universes with scale factors of the form

\begin{equation}
a^n = H^2(\eta^2 + \eta_0^2) \,,  \label{eq:bounce}
\end{equation}
where $n$ is an integer, $H$ is a constant with  dimension of 
inverse of time, and
$\eta_0$ is also a constant but with dimension of time.
Although these models are asymptotically flat in the past and in the
future, the scale factor does not approach a constant, in contrast to
the models in Sect.~\ref{sec:bounce1}.

\subsubsection{Bound Charged Particles}

\label{sec:bcp}

Here we take $n=2$ in Eq.~(\ref{eq:bounce}). In this case, the
 mean squared velocity expression is,
\begin{eqnarray}\label{eq4}
\langle\Delta u(\eta, r_1)\Delta u(\eta, r_2)\rangle =
\frac{q^2}{m^2} \int_{-\infty}^{+\infty} d\eta_2
\int_{-\infty}^{+\infty} d\eta_1 \left(\frac{1}{H^2\eta_1^2 +
H^2\eta_0^2}\right)  \\
\nonumber \times\left(\frac{1} {H^2\eta_2^2 + H^2\eta_0^2}\right)
\left[\frac{-(\eta_2-\eta_1)^2 -  r^2}{\pi^2[-(\eta_2-\eta_1)^2 +
 r^2]^3}\right].
\end{eqnarray}
This integral is evaluated in the Appendix, resulting in
Eq.~(\ref{eqdeltau10}). 
If we take the $r \rightarrow 0$ limit of this expression and use the
fact that at 
$\eta =\eta_f \gg \eta_0$, the scale factor is 
$a(\eta_f) \approx H \eta_f$, we find
\begin{equation}
\langle\Delta v^2\rangle =
\frac{q^2}{m^2}\frac{3\eta_f^2}{16H^2\eta_0^6}.
\end{equation}
The Ricci scalar curvature when $\eta =0 $ is
\begin{equation}
R_0 = -\frac{6}{H^2\eta_0^4}\,.
\end{equation}
In terms of this curvature and the redshift, 
$1+z \approx z = a_f/a(0) = \eta_f/\eta_0$,
when $\eta_f \gg \eta_0$ the mean squared velocity turns into
\begin{equation}\label{eqbounf}
\langle\Delta v^2\rangle =
\frac{q^2}{32m^2}|R_0|\, z^2.
\end{equation}
The effective temperature  is,
\begin{equation}\label{tb2}
T_{eff} \simeq
\frac{10^{-1}q^2}{k_B\lambda_c}\left(\frac{\lambda_c}{l_c}\right)^2z^2.
\end{equation}
Comparing Eq.~(\ref{tb2}) with Eq.~(\ref{tb1}) we see that they are
the same except for a numerical factor.

\subsubsection{Free Magnetic Particle}

Consider now the scale factor given by setting $n=1$ in
Eq.~(\ref{eq:bounce}). We find that the mean squared velocity is,
\begin{eqnarray}
&&\langle\Delta u(r_1,\eta)\Delta u(r_2,\eta)\rangle =
\frac{\mu^2}{a_f^4m^2}\times
\\ \nonumber &&\int^{\infty}_{-\infty}d\eta_1\int^{\infty}_{-\infty}d\eta_2
\left(\frac{1}{H^2\eta_1^2 + H^2\eta_0^2}\right)
\left(\frac{1}{H^2\eta_2^2 + H^2\eta_0^2}\right) \times
\\ \nonumber 
&&\left\{\frac{-2}{[-(\eta_2 - \eta_1)^2 + r^2]^3} - \frac{6(\eta_2 -
\eta_1)^2}{[-(\eta_2 -\eta_1)^2 + r^2]}\right\}.
\end{eqnarray}
In the coincidence limit
\begin{equation}
\langle \Delta u^2 \rangle = \frac{6\times
10^{-2}\mu^2}{a_f^4m^2H^4\eta_0^8},
\end{equation}
and the physical velocity is,
\begin{equation}
\langle \Delta v^2\rangle = \frac{
10^{-3}\mu^2|R_0|^{2}}{m^2}\, z^{-2} ,
\end{equation}
with $R_0 = {-12}/(H^4\eta_0^6)$. This is essentially the same result 
as in Eq.~(\ref{eq:fmd1}).

\subsubsection{Bound Magnetic Dipoles}

Consider now the scale factor obtained by setting $n=3$ in Eq.~(\ref{eq:bounce}).
In this case, the mean squared velocity is
\begin{eqnarray}
&&\langle \Delta u(r_1,\eta)\Delta u(r_2,\eta) \rangle =
\frac{\mu^2}{m^2}\times
\\ \nonumber &&\int^{\infty}_{-\infty}d\eta_1\int^{\infty}_{-\infty}d\eta_2
\left(\frac{1}{H^2\eta_1^2 + H^2\eta_0^2}\right)
\left(\frac{1}{H^2\eta_2^2 + H^2\eta_0^2}\right) \times
\\ \nonumber &&
\left\{\frac{-2}{[-(\eta_2 - \eta_1)^2 + r^2]^3} - \frac{6(\eta_2 -
\eta_1)^2}{[-(\eta_2 -\eta_1)^2 + r^2}\right\}.
\end{eqnarray}
In the coincidence limit

\begin{equation}
\langle \Delta u^2 \rangle = \frac{6\times
10^{-2}\mu^2}{m^2H^4\eta_0^8},
\end{equation}
with the physical velocity now given by
\begin{equation}
\langle \Delta v^2\rangle = \frac{3 \times
10^{-2}\mu^2|R_0|^{2}}{8m^2}\, z^2 \, ,
\end{equation}
where here
\begin{equation}
R_0 = -\frac{4}{H^{4/3}\eta_0^{10/3}}\, , 
\end{equation}
and $z = a(\eta_f)/a(0) = (\eta_f/\eta_0)^{2/3}$. This is of the 
same form as in Eq.~(\ref{eq:bmd1}).

\subsubsection{Free Polarizable Particles}

Now let us take  the same scale factor as in the previous subsection,
 Eq.~(\ref{eq:bounce}) with $n=3$. In this case, the 
velocity-velocity correlation
function is
\begin{eqnarray}
&&\langle \Delta u(r_1, \eta) \Delta u(r_2, \eta) \rangle =
\frac{\alpha^2}{4m^2a_f^4\pi^4} \times \\ \nonumber
&&\int_{-\infty}^{+\infty}d\eta_1\int_{-\infty}^{+\infty}d\eta_2
\frac{1}{(H^2\eta_1^2 + H^2\eta_0^2)}\frac{1}{(H^2\eta_2^2 +
H^2\eta_0^2)}\times \\ \nonumber &&\left\{ \frac{32}{[-(\eta_2 -
\eta_1)^2 + r^2]^5} + \frac{136(\eta_2-\eta_1)^2}{[-(\eta_2 -
\eta_1)^2 + r^2]^6} + \frac{144(\eta_2 - \eta_1)^4}{[-(\eta_2 -
\eta_1)^2 + r^2]^7} \right\}.
\end{eqnarray}
Doing the two integrations and taking the limit $r \rightarrow 0$,
the mean squared coordinate velocity is
\begin{equation}
\langle \Delta u^2 \rangle = \frac{3 \times
10^{-2}\alpha^2}{4m^2a_f^4\pi^2H^4\eta_0^{12}},
\end{equation}
and the proper velocity is nearly the same as in Eq.(\ref{10}),
\begin{equation}
\langle \Delta v^2 \rangle = \frac{3 \times
10^{-2}\alpha^2}{256m^2\pi^2}|R_0|^4 \, z^{-2},
\end{equation}
where now
\begin{equation}
 R_0 =- \frac{4}{H^{4/3}\eta_0^{10/3}}.
\end{equation}

\subsection{Oscillatory Expansion}

\label{sec:osc}

In this section we  add an oscillatory term in the
bouncing electric particle case studied in the last section. This
is an oscillatory expansion universe which was treated in some
works as, for example in \cite{ti86}. Here we will see the effects
of the amplitude of the oscillation in the mean squared velocity
for the charged particle case. 
The scale factor is taken to be
\begin{equation}
a^2=a_0^2(\eta^2 + \eta_0^2) + a_1^2\cos(\omega\eta),
\label{eq:osc-scale}
\end{equation}
where $a_0$ and the frequency $\omega$ are constants with dimension of
the inverse of time, 
the amplitude $a_1$ is a dimensionless constant, and 
$\eta_0$ is a constant  with dimension of time. 
Thus the oscillatory term is a small perturbation of the case studied
in Sect.~\ref{sec:bcp}. 
The mean squared velocity is
\begin{eqnarray}\label{eqos1}
\langle\Delta u(\eta, r_1)\Delta u(\eta, r_2)\rangle =
\frac{q^2}{m^2}\int^{+\infty}_{-\infty}d\eta_2\int^{+\infty}_{-\infty}d\eta_1
\left[ \frac{-(\eta_2-\eta_1)^2 - r^{'2}}{\pi^2[-(\eta_2-\eta_1)^2
+ r^2]^3}
\right]\\
\nonumber  \times\left(\frac{1}{a_0^2\eta_1^2 + a_0^2\eta_0^2 +
a_1^2\cos(\omega\eta_1)}\right)\left(\frac{1}{a_0^2\eta_2^2 +
a_0^2\eta_0^2 + a_1^2\cos(\omega\eta_2)}\right).
\end{eqnarray}

In Appendix B this integral is evaluated, and can be shown to lead to the result
\begin{equation}\label{eqosc}
\langle\Delta v(r_1) \Delta v(r_2)\rangle = 
\langle\Delta v^2 \rangle_0 + \frac{q^2 a_1^2\eta_f^6}{m^2 \eta_0^4
  l_f^4}
\left(3\omega\eta_0\sinh(\omega\eta_0)+ \cosh(\omega\eta_0) \right)\, ,
\end{equation}
where $l_f = a_f r$, and $\langle\Delta v^2 \rangle_0$ is the $a_1=0$
result found in 
Eq.(\ref{eqbounf}).
Notice that the velocity-velocity correlation function in this model
will diverge in the 
coincidence limit. So, we can not obtain the mean squared velocity in
this limit. This 
reflects a breakdown of our model in which the particles are treated 
as classical point
objects. Our model requires that $l_f \gg \lambda_C$, where
$\lambda_C$ is the electron 
Compton wavelength. In any case, our perturbative result requires that
the second term in 
Eq.~(\ref{eqosc}) be small compared to $\langle\Delta v^2 \rangle_0$. 
Nonetheless, we
can conclude that the oscillations tend to increase the mean squared
velocity in a way that 
grow exponentially with $\omega$ in the limit that $\omega \eta_0 \gg 1$.

\subsection{de Sitter Space}

In this section, we will investigate the effects of the vacuum
fluctuations in de Sitter 
space-time. It is well known that de Sitter space can be  considered
as some special stage of the universe history, which is 
known as the inflationary
phase of the universe. It was Guth \cite{g81} who first 
noticed that using some exponential
expansion of the universe 
it would be possible to solve  three of the standard universe's 
modelproblems: 1) the flatness 
problem, 2) the horizon problem and 3) the primordial
monopoleproblem. 
This scenario was 
extensively developed since the Guth's original work (see some good
reviews about inflation in \cite{linde07}, \cite{bornerb00k03}) and n
owadays it seems to be in
good agreement with the observations \cite{llbook00},
\cite{spergeletal06}.

We  use  a scale factor in the form:
\begin{equation}\label{dsscale}
 a = -\frac{1}{H\eta},
\end{equation}
where $ -\infty < \eta < 0$. 
We restrict our attention to the range $\eta_i \leq \eta \leq \eta_f$, 
 where $|\eta_f| \ll |\eta_i|$.

\subsubsection{Bound Charged Particle}

Here the velocity-velocity correlation function is given by,
\begin{eqnarray}
\langle\Delta u(\eta, r_1)\Delta u(\eta, r_2)\rangle =
\frac{q^2}{m^2}\int_{\eta_i}^{0}d\eta_1\int_{
\eta_i}^{0}d\eta_2(H\eta_1)^2(H\eta_2)^2
\\ \nonumber \times\left[\frac{-(\eta_2-\eta_1)^2-
r'^2}{\pi^2[-(\eta_2-\eta_1)^2 +  r^2 ]^3}\right].
\end{eqnarray}
Notice that the upper value in the integral range is put to be 
zero to simplify our calculations, but in fact it is very small but not null.
Using Maple, we find  the following result,
\begin{equation}
\langle\Delta u(r_1) \Delta u(r_2)\rangle = \frac{q^2H^4}{m^2
r^2}\left[3\eta_i^4 - 2\eta_i^2 r'^2\right].
\end{equation}If $|\eta_i| \gg  r$ the coordinate velocity is:
\begin{equation}
\langle\Delta u(r_1)\Delta u(r_2)\rangle =
\frac{q^2H^4\eta_i^4}{m^2 r^2},
\end{equation}
and consequently the physical velocity is,
\begin{equation}\label{eqds}
\langle\Delta v(r_1)\Delta v(r_2)\rangle = \frac{q^2}{m^2 l_i^2}.
\end{equation}
Our answer is positive and constant and as in the oscillatory case 
it depends on the initial proper particle separation, $l_i$. 
If $l_i \approx \lambda_c$, then Eq.~(\ref{eqds}) reduces to,
\begin{equation}
\langle\Delta v(r_1)\Delta v(r_2) \rangle \simeq \frac{q^2}{4}
\simeq 10^{-2}.
\end{equation}

\subsubsection{Free Magnetic Dipoles}

Using (\ref{dsscale}) and (\ref{eqmag2}) we obtain the following 
mean squared velocity,
\begin{eqnarray}
\langle \Delta u(r_1,\eta)\Delta u(r_2,\eta) \rangle =
&&\frac{\mu^2}{a_f^4m^2}
\int_{-\eta_i}^{0}d\eta_1\int_{-\eta_i}^{0}d\eta_2
\left(H\eta_2\right) \left(H\eta_1\right) \times
\\ \nonumber &&
\left\{\frac{-2}{[-(\eta_2 - \eta_1)^2 + r^2]^3} - \frac{6(\eta_2 -
\eta_1)^2}{[-(\eta_2 -\eta_1)^2 + r^2]^4}\right\},
\end{eqnarray}
which, after integrations results in,
\begin{equation}
\langle \Delta u(r_1,\eta)\Delta u(r_2,\eta) \rangle =
\frac{\mu^2H^2}{6a_f^4m^2r^2}.
\end{equation}
{}From the equation above it is not possible to get the coincidence 
limit because the $r^2$-divergence. But we know that $l_f = a_f r$, then,
\begin{equation}\label{dsf1}
\langle \Delta v(r_1,\eta)\Delta v(r_2,\eta) \rangle =
\frac{\mu^2H^2}{6m^2l_f^2}.
\end{equation}
Considering $\mu \sim q/m$,  $m \sim 1/\lambda_c$ and
$l_f \sim \lambda_c$, we have
\begin{equation}
\langle \Delta v(r_1,\eta)\Delta v(r_2,\eta) \rangle \simeq
\frac{q^2H^2\lambda_c^2}{6}.
\end{equation}
The effect in de Sitter universe does not depend of the time. 
We could also make a estimate of the constant $H$, which has an 
inverse of length dimension . If $H \sim 1/l$, we have,
\begin{equation}\label{dsf2}
\langle \Delta v(r_1,\eta)\Delta v(r_2,\eta) \rangle \simeq
\frac{q^2\lambda_c^2}{6l^2}.
\end{equation}
Assuming that $l \sim \lambda_c$, we get
\begin{equation}\label{dsf4}
\langle \Delta v(r_1,\eta)\Delta v(r_2,\eta) \rangle \simeq
\frac{q^2}{6},
\end{equation}
which is basically the same result we found for the bound charge case.

\subsubsection{Bound Magnetic Dipoles}

\label{sec:deSbmd}

Using the scale factor given by Eq.~(\ref{dsscale}) and 
Eq.~(\ref{eq2mag2}),  we find that the mean squared velocity is given by
\begin{eqnarray}
\langle \Delta u(r_1,\eta)\Delta u(r_2,\eta) \rangle =
&&-\frac{\mu^2}{m^2}
\int_{\eta_i}^{0}d\eta_1\int_{\eta_i}^{0}d\eta_2
\left(H\eta_2\right)^3 \left(H\eta_1\right)^3 \times
\\ \nonumber &&
\left\{\frac{2}{[-(\eta_2 - \eta_1)^2 + r^2]^3} + \frac{6(\eta_2 -
\eta_1)^2}{[-(\eta_2 -\eta_1)^2 + r^2]^4}\right\}.
\end{eqnarray}
In the limit which $r$ is very small, we have
\begin{equation}
\langle \Delta u(r_1,\eta)\Delta u(r_2,\eta) \rangle =
-\frac{\mu^2}{m^2}\frac{\eta_i^4H^6}{2r^2}.
\end{equation}
Thus, the physical velocity-velocity correlation function is given by
\begin{equation}
\langle \Delta v(r_1,\eta)\Delta v(r_2,\eta) \rangle =
-\frac{q^2H^2\lambda_c^2}{2},
\end{equation}
which is negative.
Making the same estimates as in the free magnetic particle case for
$H$, we obtain,
\begin{equation}\label{dsf3}
\langle \Delta v(r_1,\eta)\Delta v(r_2,\eta) \rangle =
-\frac{q^2\lambda_c^2}{2l^2}.
\end{equation}

Negative mean squared velocities have been found by previous 
authors~\cite{wkf02,yf04,sw07}, and can be interpreted as a reduction 
in quantum uncertainty. It is well known that a quantum massive
particle 
is described by a wave packet which must have a position and momentum 
uncertainty given by the Heisenberg uncertainty principle 
$\langle \Delta p_x \Delta x \rangle \geq 1$. If the uncertainty in
position is such that $\Delta x \lesssim l$, where $l$ is the 
average separation between two dipoles, we have, 
$\langle \Delta v_x \rangle_q \gtrsim 1/(lm)$, or
\begin{equation}\label{qds3}
\langle \Delta v_x^2 \rangle_q \gtrsim \frac{\lambda_c^2}{l^2}\,,
\end{equation}
which is larger than the magnitude of the right-hand side of Eq.~(\ref{dsf3}).

\subsection{Radiation Dominated Universe}

The radiation dominated era of our universe is usually defined as
the early period when radiation and relativistic particles were
usually more important than ordinary matter. Here we will
investigate the effects of this important universe stage in the
average squared velocity and evaluate the effective particle's
temperature in the beginning of this era in the bounded electric
particle case.
Consider the scale factor,
\begin{equation}\label{eq14}
a^2 = H^2\eta^2\,.
\end{equation}
Then the mean squared velocity is given by
\begin{eqnarray}
\langle\Delta u(\eta, r_1) \Delta u(\eta, r_2)\rangle =
\frac{q^2}{m^2}\int_{\eta_0}^{\infty}d\eta_1\int_{\eta_0}^{\infty}d\eta_2
\left(\frac{1}{H^2\eta_1^2}\right)\left(\frac{1}{H^2\eta_2^2}\right)
\\ \nonumber \times\left[\frac{-(\eta_2-\eta_1)^2-
r'^2}{\pi^2[-(\eta_2-\eta_1)^2 +  r^2 ]^3}\right].
\end{eqnarray}
Taking into account the condition $r \ll \eta_0$, this integral results in
\begin{equation}\label{eqcorrec1}
\langle\Delta u^2 \rangle = \frac{C_1q^2}{m^2\eta_0^6H^4},
\end{equation}
where $C_1 \simeq 6\times10^{-5}$.
The physical velocity can be written as
\begin{equation}
\langle\Delta v^2\rangle =
\frac{C_1q^2}{m^2}\frac{\eta_f^2}{H^2\eta_0^6}.
\end{equation}
The redshift factor here is $1+z \approx z = \eta_f/\eta_0$. 
The scalar curvature
vanishes for this metric, but a reasonable measure of the 
characteristic curvature is a typical component of the Ricci tensor 
in an orthonormal frame, which gives
\begin{equation}
R_0 \approx \frac{1}{H^2 \eta_0^4} \,.
\end{equation}
Thus, we can write
\begin{equation}
\langle\Delta v^2\rangle =\frac{C_1q^2}{m^2}\, R_0\, z^2 \, ,
\end{equation}
which is essentially the same as the result found in previous cases in
 Eqs.~(\ref{eq:bcp1}) and (\ref{eqbounf}).

\subsection{Matter Dominated Universe}

Now consider bound charged particles and a scale factor of the form,
\begin{equation}\label{eq19}
a^2 = H^4\eta^4.
\end{equation}
The mean squared velocity in this case is
\begin{eqnarray}
\langle\Delta u(\eta, r_1)\Delta u(\eta, r_2)\rangle =
\frac{q^2}{m^2}\int_{\eta_0}^{\infty}d\eta_1\int_{\eta_0}^{\infty}d\eta_2
\left(\frac{1}{H^4\eta_1^4}\right)\left(\frac{1}{H^4\eta_2^4}\right)
\\ \nonumber \times\left[\frac{-(\eta_2-\eta_1)^2-
r'^2}{\pi^2[-(\eta_2-\eta_1)^2 +  r^2 ]^3}\right].
\end{eqnarray}

Performing the integrals in the limit $r \ll \eta_0$, we obtain the
following result
\begin{equation}
\langle\Delta u^2\rangle = \frac{10^{-5}q^2}{m^2\eta_0^{10}H^8}.
\end{equation}
The physical velocity  is given by
\begin{equation}
\langle\Delta v^2\rangle =
\frac{C_2q^2}{m^2}\frac{\eta_f^4}{H^4\eta_0^{10}} =
 \frac{C_2q^2}{m^2}\, R_0 \, z^2 \, ,
\end{equation}
where $C_2 \simeq 10^{-5}$, $z=(\eta_f/\eta_i)^2$, and $R_0 = 1/(H^4 \eta_0^6)$.
Again this is of the same form as Eq.~(\ref{eq:bcp1}).

\section{Summary and Discussion}

\label{sec:final}

We have investigated the Brownian motion  of particles coupled to the
 electromagnetic vacuum fluctuation in Robertson-Walker universes. We 
considered several types of particles, including ones with electric
charge, 
a magnetic dipole moment, and electric
polarizability. We also allowed both the possibility that the
particles are 
free, moving apart on the average as the universe expands, or bound by
a force which cancels the effect of the expansion. Our results for the
mean squared velocity induced by
quantum fluctuations can typically be written in terms of a
characteristic 
measure of the space-time curvature, $R_0$, and a redshift factor
$z$. Our treatment assumes semiclassical point particles which always
move non-relativistically in the comoving
frame. Thus our results are restricted to cases where  
$\langle\Delta v^2\rangle \ll 1$. We should also note that we are 
working in a regime where quantum particle creation by the
gravitational field~\cite{Parker69} is small, so we should also 
require that $R_0/m^2 \ll 1$. 

In many cases, such as the classes of scale factors studied in 
Sects.~\ref{sec:bounce1} and \ref{sec:bounce2}, the effect for bound 
particles tends to grow with increasing redshift factor, where as that 
for free particles goes the other way. This might
be due to the fact that the bound particles are not on the average 
moving on geodesics, and be subject to an acceleration radiation
effect of the type first studied by Unruh~\cite{Unruh76}. However, 
in other case , such as those treated in
Sect.~\ref{sec:bounded-expand}, 
the effect scales differently with redshift. These results need to be 
better understood. 

In most of the cases studied, the mean squared velocity is finite and 
positive. However, in a few cases, such as in Sects.~\ref{sec:osc} and 
\ref{sec:deSbmd} , we found a velocity-velocity correlation function 
which is singular at spatially coincident points and can be negative. 
Both of these phenomena signal a breakdown of our approximation
of point classical particles coupled to a quantized field.

However, our view is that these results still have physical content 
if properly interpreted. The spatial separation of particles should 
always be large compared to the Compton wavelength, and the separation 
should be sufficiently large to insure that
$\langle\Delta v_1 \Delta v_2 \rangle \ll 1$. With these restrictions, 
one can still conclude from Eq.~(\ref {eqosc}) that oscillations 
superimposed upon a uniform bouncing universe, as in 
Eq.~(\ref{eq:osc-scale}), lead to additional heating. Similarly, 
the cases where $\langle\Delta v^2\rangle < 0$ signal a reduction in quantum
uncertainty, or a form of gravitational squeezing, analogous to
 effects near mirrors discussed in Refs.~\cite{wkf02,yf04,sw07}.

One of the motivations of this study is theoretical, to better 
understand quantum Brownian motion in a curved spacetime, as an analog 
model for the effects of the quantum fluctuations of gravity. However, 
it is also natural to enquire as to whether our 
results could have application to realistic cosmological models.

One possibility is an additional reheating mechanism after the end of
 inflation. If inflation ends quickly, it is likely that the reheating 
temperature will exceed the effective temperature due to Brownian
motion.
 If reheating is inefficient, however, 
there is a possibility that Brownian motion could play a role.

Recall that the results in this paper are restricted to the case on 
non-relativistic motion, or when the temperature is small compared to 
the particle's rest mass energy. This severely limit the use of these 
result for electrons or nucleons. The restriction
is less severe for very massive particles, such as  
``wimpzillas''~\cite{wimp,wimp2}. These are hypothetical particles 
with masses up to the Planck scale produced at the end of inflation
by, for example, gravitational particle  
creation~\cite{Parker69,Ford87}. We plan to extend the study in 
the present paper to the relativistic motion case, which will lift 
this  restriction, and to give a more detailed discussion of applications
to inflationary cosmology. Another possible extension is to the case
of Brownian motion produced by fluctuation of non-Abelian gauge fields.

\section*{Acknowledgements}

CHGB is in debt to Physics and Astronomy Department at Tufts
University for their hospitality and to Coordena\c{c}\~ao de
Aperfei\c{c}oamento de Pessoal de N\'{\i}vel Superior (CAPES) of
Brazil for financial support. VBB thanks Conselho Nacional de
Desenvolvimento Cient\'{\i}\-fi\-co e Tecnol\'ogico (CNPq),
FAPESQ-PB/CNPq(PRONEX) and FAPES-ES/CNPq(PRONEX) of Brazil for
financial support. LHF thanks the National Science Foundation for
support under Grant PHY-0555754.

\appendix

\section{Minkowski Space Correlation Functions}

\label{app:field}

Here we will briefly summarize the calculation of the components 
of the electromagnetic field strength tensor correlation function 
in flat spacetime. Write the Minkowski metric in the form

\begin{equation}
ds^2 = -d\eta^2 + dx^2 + dy^2 + dz^2 \,. \label{eq: FSTmetric}
\end{equation} 
We are interesting in the components of the field strength tensor 
correlation function,
\begin{equation}
\langle F_{\mu\nu}(x)\, F_{\alpha\beta}(x') \rangle_M \,.
\end{equation}
These are easily computed from the vector potential correlation 
function using the relation 
$F_{\mu\nu}= \partial_\mu A_\nu - \partial_\nu A_\mu$. 
The vector potential correlation function can, in a suitable gauge, 
be written as
\begin{equation}
\langle A_{\mu}(x)\, A_{\nu}(x') \rangle_M = \frac{\eta_{\mu\nu}}{8 \pi^2\,
  \sigma}\,,
\end{equation}
where $\eta_{\mu\nu}$ is the Minkowski metric tensor, and
\begin{equation}
\sigma = \frac{1}{2} [-(\eta-\eta')^2 +(x-x')^2 +(y-y')^2 +(z-z')^2] .
\end{equation}

The electric field correlation function, for example, is

\begin{equation}
\langle E^x(x)\, E^x(x') \rangle =
\langle F^{\eta x}(x)\, F^{\eta x}(x') \rangle = 
 -\frac{\Delta \eta^2 + r'^2}{\pi^2[-\Delta \eta^2 + r^2]^3} \, ,
\label{eq:ExEx}
\end{equation}
where $r'^2 = r^2 -2 \Delta x^2$ and $r^2 =\Delta x^2+ \Delta y^2 + \Delta z^2$.


\section{Evaluation of Integrals}

\label{app:integrals}

\subsection{Evaluation of Eq.~(\ref{eq5})}

We may evaluate the two integrals in Eq. (\ref{eq5}) using the residue 
theorem. First, we evaluate the $\eta_1$ integral indicated by:
\begin{equation}\label{b1eq5}
I_1 = \int_{-\infty}^{+\infty}d\eta_1
\left(\frac{\eta_1^2+G^2\eta_0^2}{\eta_1^2+\eta_0^2}\right)^2
 \left[\frac{-(\eta_2-\eta_1)^2 - r^{'2}}{\pi^2[-(\eta_2-\eta_1)^2
+  r^2]^3}\right],
\end{equation}

\begin{figure}
\begin{center}
\includegraphics[width=0.50\textwidth]{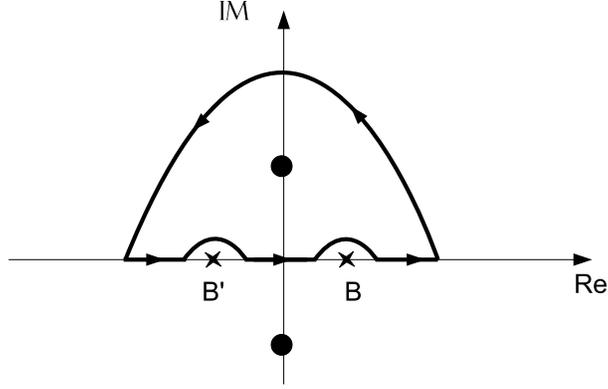}
\end{center}
\caption{Integration contour for Eq. (\ref{b1eq5}). The lower order
  poles, which are a pure  imaginary number, are indicated by the d
ot symbol $\bullet$. The higher order poles has only a real part, 
as indicated by the $\textbf{X}$ symbol and their contribution is 
a pure imaginary number. } \label{figb1}
\end{figure}

\begin{figure}
\begin{center}
\includegraphics[width=0.35\textwidth]{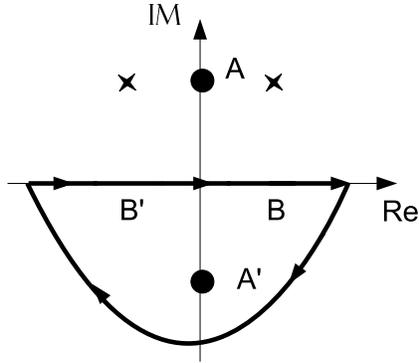}
\end{center}
\caption{This figure illustrated the easiest integration contour 
we may choose to evaluate Eq. (\ref{eqb2}). It is easiest because 
we avoid the highest order poles represented by the symbol 
$\textbf{X}$ in the figure. So the contribution of $I_2$ integral 
is only due to the negative and imaginary lower order pole which
is indicated by the dot symbol $\bullet$ in the lower half plane. }
\label{figb2}
\end{figure}

In order to do that, let us choose a contour that avoids the 
third order poles located in the real axis at 
$\eta_1 = \pm  r + \eta_2$ (as indicated, in Fig.~(\ref{figb1}), 
by the letters $B$ and $B^{'}$). Then, we evaluate the integral 
considering only one of
the second order poles, namely, $\eta_1 = \pm i\eta_0$. 
In Fig.~(\ref{figb1}) we illustrate only the contour in the upper 
half plane. The contour in the lower half plane would give us the 
same answer, according to the residue theorem. So the first integral
$(I_1)$ has the form

\begin{eqnarray}
I_1 =  \nonumber \frac{\eta_0(G^2-1)^2[-(\eta_2-i\eta_0)^2 -
r^2]}{2\pi [-(\eta_2-i\eta_0)^2+ r^2]^3} +
\frac{2\eta_0(G^2-1)[-(\eta_2-i\eta_0)^2-
r^2]}{\pi^2[-(\eta_2-i\eta_0)^2+ r^2]^3}
\\ \nonumber
-\frac{3\eta_0^2i(\eta_2-i\eta_0)(G^2-1)^2}{\pi[-(\eta_2-i\eta_0)^2+
r^2]^3} + \frac{3\eta_0i(G^2-1)(\eta_2-i\eta_0)[-(\eta_2 -
i\eta_0)^2 -  r^2] }{\pi [-(\eta_2-i\eta_0)^2- r^2]^4}.
\end{eqnarray}

Now the second integral is
\begin{equation}\label{eqb2}
I_2 =
\int^{+\infty}_{-\infty}d\eta_2\left(\frac{\eta_2^2+G^2\eta_0^2}
{\eta_2^2+\eta_0^2}\right)^2I_1.
\end{equation}

The integrand in $I_2$ has a second order pole located at 
$\eta_2 = \pm i\eta_0$ and three third and one fourth orders poles at 
$\eta_2 = \pm  i\eta_0$. We evaluate $I_2$ in the lower half plane 
to avoid the highest order poles, as indicated in Fig.~(\ref{figb2}). 
Even if we had chosen a contour in the upper half plane, our answer 
would be the same, but the way to do that would be harder. So, the 
final result for the mean squared velocity 
($\langle \Delta u(r_1)\Delta u(r_2) \rangle = \frac{q^2}{m^2}I_1I_2)$ is,

\begin{eqnarray}
\langle\Delta u(r_1) \Delta u(r_2)\rangle =
\frac{q^2\eta_0\Theta^2}{m^2}\left\{\frac{\eta_0\Theta^2(12\eta_0^2-
r'^2)}{(4\eta_0^2+ r^2)^3} +
\frac{\eta_0\Theta(12\eta_0^2- r'^2)}{(4\eta_0^2+
r^2)^3} \right\}
\\ \nonumber + \frac{q^2\eta_0\Theta^2}{m^2}
\left\{\frac{-3\eta_0^3\Theta^2}{(4\eta_0^2+
r^2)^3} + \frac{3\eta_0^3\Theta(12\eta_0^2-
r'^2)}{(4\eta_0^2+r^2)^4} \right\} \\ \nonumber +
\frac{4q^2\eta_0\Theta}{m^2}\left\{\frac{\eta_0\Theta^2(12\eta_0^2-
r'^2)}{(4\eta_0^2+ r^2)^3} +
\frac{\eta_0\Theta(12\eta_0^2- r'^2)}{(4\eta_0^2+
r^2)^3} \right\}
\\ \nonumber +
\frac{4q^2\eta_0\Theta}{m^2}\left\{\frac{-3\eta_0^3\Theta^2}{(4\eta_0^2+
r^2)^3} + \frac{3\eta_0^3\Theta^2(12\eta_0^2-
r'^2)}{(4\eta_0^2+ r^2)^4} \right\} \\ \nonumber
-\frac{q^2\eta_0^2\Theta^2}{m^2}\left\{\frac{3\eta_0^2\Theta^2}
{2(4\eta_0^2+
r^2)^3} - \frac{3\eta_0^2\Theta(12\eta_0^2-
r'^2)}{(4\eta_0^2+ r^2)^4} \right\}  \\ \nonumber
-\frac{q^2\eta_0^2\Theta^2}{m^2}\left\{\frac{12\eta_0^2\Theta}{(4\eta_0^2+
r^2)^3} + \frac{12\eta_0^2\Theta(12\eta_0^2-
r'^2)}{(4\eta_0^2+r^2)^4} \right\} \\ \nonumber -
\frac{q^2\eta_0^2\Theta^2}{m^2}\left\{\frac{72\eta_0^4\Theta^2}
{(4\eta_0^2+
r^2)^4} - \frac{48\eta_0^4\Theta(12\eta_0^2-
r'^2)}{(4\eta_0^2+ r^2)^5} \right\}&, \label{eq:delu1}
\end{eqnarray}
where $\Theta \equiv G^2 - 1$.

\subsection{Evaluation of Eq.~(\ref{eq667})}

Next we turn to the evaluation of Eq~(\ref{eq667}). 
Define a new $I_1$ integral as

\begin{equation}\label{eqsm1}
I_1 =  \int_{-\infty}^{+\infty} d\eta_1 \left(\frac{1}{a_0^2 +
a_1^2\tanh\left(\frac{\eta_1}{\eta_0}\right)}\right)
\left[\frac{-(\eta_2-\eta_1)^2 - r^{'2}}{\pi^2[-(\eta_2-\eta_1)^2
+ r^2]^3}\right]\,.
\end{equation}
We see that in the $\eta_1$ integral we have third order poles 
located in the real axis at: $\eta_1 = \pm r +\eta_2$, and an infinite 
number of poles at $\tanh(\eta_1/\eta_0) = -\alpha^2$, where 
$\alpha^2 \equiv a_0^2/a_1^2 > 1$. These are the first order
poles, and  can be written as

\begin{equation}
\eta_{1k} = \eta_{10} + \frac{1}{2}(2k+1)i\pi\eta_0,
\end{equation}
with $k  = 0, \pm 1, \pm 2, ...$, and 
$\eta_{10} = -\frac{\pi i}{2}\eta_0 -
\frac{1}{2}\eta_0\ln\left(\frac{\alpha^2 
+1}{\alpha^2 - 1}\right)$.

\begin{figure}
\begin{center}
\includegraphics[width=0.50\textwidth]{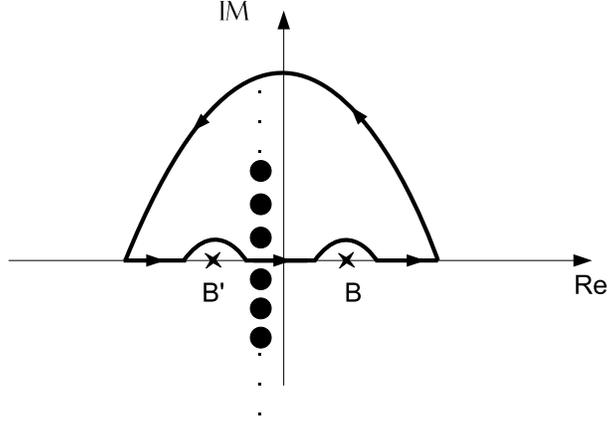}
\end{center}
\caption{One possible contour to integrate Eq. (\ref{eqsm1}). 
We avoid the higher order poles, indicated by $\textbf{X}$ symbol, 
because their contribution is a pure imaginary number. The contour is, 
however, an infinity contour that encloses the infinity
positives lower order poles which are indicated by the symbol 
$\bullet$. }\label{figsm1}
\end{figure}

Following the contour indicated in  Fig. (\ref{figsm1}) we obtain,

\begin{equation}
I_1 =
\sum_{k=1}^{\infty}\frac{2i\eta_0\cosh^2
\left(\frac{\eta_{1k}}{\eta_0}\right)}{a_1^2}\left[\frac{-(\eta_2
- \eta_{1k})^2 -  r^{'2}}{\pi [-(\eta_2 - \eta_{1k})^2 +
r^2]^3}\right],
\end{equation}
Note that a contour in the lower half plane would give us the same
answer. With this result, the $\eta_2$ integral turns into

\begin{eqnarray}\label{eqsm2}
I_2 =
\sum_{k=1}^{\infty}\frac{2i\eta_0}{a_1^2}\cosh^2\left(\frac{\eta_{1k}}
{\eta_0}\right)\int_{-\infty}^{+\infty}\frac{d\eta_2}{a_0^2
+ a_1^2\tanh\left(\frac{\eta_2}{\eta_0}\right) } \\ \nonumber
\times\left[\frac{-(\eta_2 - \eta_{1k})^2 - r^{'2}}{\pi [-(\eta_2
- \eta_{1k})^2 +  r^2]^3}\right].
\end{eqnarray}

\begin{figure}
\begin{center}
\includegraphics[width=0.35\textwidth]{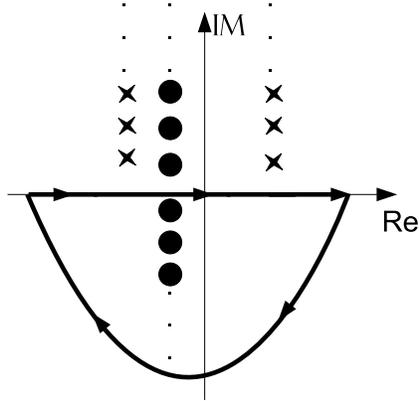}
\end{center}
\caption{The figure shows the easiest possible contour to integrate 
Eq.~(\ref{eqsm2}), because the infinities higher order poles 
(the {$\textbf{X}$} symbol) are excluded. The enclosed poles 
are the infinities lower orders, which are indicated by $\bullet$
symbol. }\label{figsm2}
\end{figure}

It has third order poles, but now located at, 
$\eta_2 = \pm  r + \eta_{1k}$, and  first order poles at
$\eta_{2l} = \eta_{20} + \frac{1}{2}(2l + 1)i\pi\eta_0$, with 
$\eta_{20} = \eta_{10}$, and $l = 0, \pm 1, \pm 2,$ ...These
singularities are located in the complex plane as indicated in 
Fig.~(\ref{figsm2}). To avoid the third order poles we choose a 
contour in the lower half plane, however the same answer is required 
if we had chosen the contour in the upper plane.

So our result is,
\begin{equation}\label{vel2}
\langle\Delta u(r_1) \Delta u(r_2)\rangle = -4
\frac{\eta_0^2}{a_1^4}\sinh^4\left[\frac{1}{2}
\ln\left(\frac{\alpha^2+1}{\alpha^2-1}\right)\right]\; S\, , 
\end{equation}
where 

\begin{equation}
S =
\sum_{k=1}^{\infty}\sum_{l^{'}=1}^{\infty}\left[\frac{\pi^2(l^{'}
+ k)^2\eta_0^2 -  r^2 }{[\pi^2(l^{'} + k)^2\eta_0^2+
r^2]^3}\right] = \sum_{j=2}^{\infty} (j -1)\frac{\pi^2 j^2\eta_0^2
- r^2 }{[\pi^2 j^2\eta_0^2 + r^2]^3}.
\end{equation}
In the second form we transformed the double sum into a single sum 
by defining $j \equiv k + l^{'} - 1$ and $l^{'} \equiv -l$.
This sum can be evaluated explicitly to yield

\begin{eqnarray}
&S = \frac{1}{\pi^4\eta_0^4}\left[ -\frac{i\pi\eta_0}{2
r}\Psi\left(1,2-\frac{i r}{\eta_0\pi}\right)
 +
 \frac{i\pi\eta_0}{2 r}\Psi\left(2,2+\frac{i
   r}{\eta_0\pi}\right)\right] 
                                \\ \nonumber
& + \frac{1}{\pi^4\eta_0^4}\left[\left(-\frac{1}{4} -
\frac{i\pi\eta_0}{4 r}\right)\Psi\left(2,2-\frac{i
r}{\eta_0\pi}\right) + \left(-\frac{1}{4} + \frac{i\pi\eta_0}{4
r}\right)\Psi\left(2,2+\frac{i r}{\eta_0\pi}\right)\right],
\end{eqnarray}
where $\Psi(n,x)$ is the nth Polygamma function 
(This calculation was done using the algebraic program, Maple).

\subsection{Evaluation of Eq.~(\ref{eq4})}

Let the $I_1$ integral in the variable $\eta_1$. We may evaluate 
it using again the residue theorem. The $\eta_1$ integral has third
order poles at $\eta_1 = \pm r + \eta_2$, and single poles at 
$\eta_1 = \pm i\eta_0$. Here, we can use again the integration contour
in Fig.~(\ref{figb1}). However, the symbol $\bullet$
represents  singles order poles and not second order poles as in the 
case of Eq. (\ref{b1eq5}) which corresponds to the asymptotically flat 
bouncing universe. Thus, the  integral $I_1$ is expressed as,

\begin{eqnarray}
I_1 = \int_{-\infty}^{+\infty} d\eta_1 \left(\frac{1}{H^2\eta_1^2
+ H^2\eta_0^2}\right)   \left[\frac{-(\eta_2-\eta_1)^2 -
 r^2}{\pi^2[-(\eta_2-\eta_1)^2 +  r^2]^3}\right]&& \vspace{10.0cm} \\
\nonumber = \frac{-(\eta_2 - i\eta_0)^2 -  r^2}{\eta_0
H^2[-(\eta_2-i\eta_0)^2 + r^2]^3}.
\end{eqnarray}
The $\eta_2$ integral is given by

\begin{equation}\label{thisone1}
I_2 = \int^{+\infty}_{-\infty} \frac{1}{H^2\eta^2 +
H^2\eta_0^2}I_1.
\end{equation}
It has two single poles at $\eta_2 = \pm i\eta_0$ and two 
third order poles at $\eta_2 = \pm  r + i\eta_0$. In this case
 we choose a contour in the lower half plane, as in the
 Fig.~(\ref{figb2}), with the symbol $\bullet$ representing now first
 order poles. This
contour avoids the third order poles. Even if we choose the contour 
in the upper half plane, our result would be the same.
Then, after the integrations Eq. (\ref{thisone1}) results in

\begin{equation}\label{eqdeltau10}
\langle\Delta u(r_1)\Delta u(r_2)\rangle =
\frac{q^2}{m^2}\left[\frac{3\eta_0^2 -  r^2}{\eta_0^2
H^4[4\eta_0^2 +  r^2]^3}\right].
\end{equation}

\subsection{Evaluation of Eq.~(\ref{eqos1})}

\begin{figure}
\begin{center}
\includegraphics[width=0.50\textwidth]{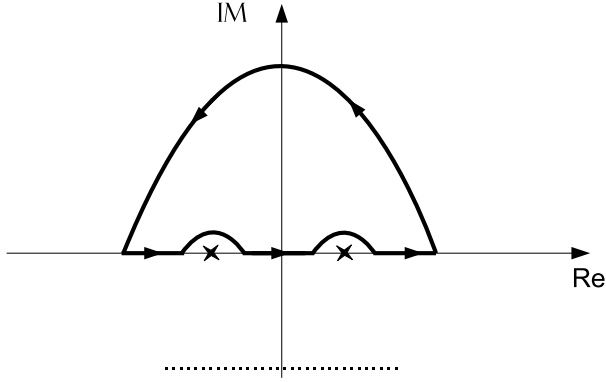}
\end{center}
\caption{This figure shows that it is possible to choose one contour 
where the integral is apparently null. }\label{figos1}
\end{figure}

Next we consider the evaluation of Eq.~(\ref{eqos1}). 
The $\eta_1$ singularities has third order poles  located at 
$\eta_1 = \pm r + \eta_2$, and first order poles at
$\cos(\omega\eta_1)= -\alpha^2\eta_1^2 - \alpha^2\eta_0^2$, where 
$\alpha^2 \equiv a_0^2/a_1^2$ or 
$\eta_{1k} = -\frac{i}{\omega}\ln\left(\alpha^2 B(\eta_{1k}) + 
\sqrt{\alpha^2 B(\eta_{1k}) -1 }\right) + 
\frac{1}{2}(2k + 1)\frac{\pi}{\omega}$,
with $k = 0, \pm 1, \pm 2,$ ..., and 
$B(\eta_{1k} \equiv \eta_1^2 + \eta_0^2)$. 
These singularities are located as indicated in Fig.~(\ref{figos1}), 
with the $\textbf{X}$ symbol now representing the third order poles 
and the several $\cdot\cdot\cdot\cdot\cdot$
symbols representing the infinities numbers of first order poles.
So apparently, this integral is zero because we do not have any poles 
enclosed by one of the possible contours. However, if $\eta = ix$, 
we can find two real poles in the integrand, because 
$\cos(\omega\eta_1) = -\alpha^2\eta_1^2 - \alpha^2\eta_0^2
\Rightarrow \cosh(\omega x) = -\alpha^2 x^2 - \alpha^2\eta_0^2$.

\begin{figure}
\begin{center}
\includegraphics[width=0.50\textwidth]{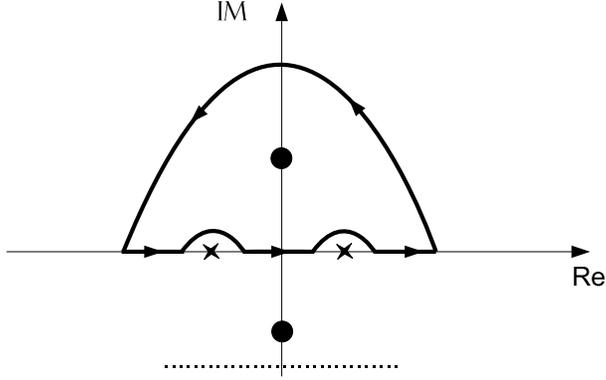}
\end{center}
\caption{This figure shows that Eq. (\ref{eqos1}) has at least one 
imaginary pole located at $\eta_0$ which can be enclosed by the
contour of Fig.~(\ref{figos1}). }\label{figos2}
\end{figure}

Now we have the picture indicated in Fig. (\ref{figos2}). Then, we 
have at least one contribution due the imaginary first order pole 
represented there by the symbol $\bullet$. Thus, the $\eta_1$ integral 
($I_1$) is,

\begin{equation}
I_1 = \frac{2}{2a_0^2x -
a_1^2\omega\sinh(x\omega)}\left[\frac{-3(\eta_2 - ix )^2 -
r^{'2}}{\pi[-(\eta_2-ix)^2+ r^2]^3}\right],
\end{equation}
and the $I_2$ is defined as:

\begin{eqnarray}\label{eqos2}
I_2 = \frac{2}{2a_0^2x -
a_1^2\omega\sinh(x\omega)}\int^{+\infty}_{-\infty}\frac{d\eta_2}{a_0^2\eta_2^2
+ a_0^2\eta_0^2 + a_1^2\cos(\omega\eta_2) }\\ \nonumber \times
\left[\frac{-3(\eta_2 - ix )^2 -  r^{'2}}{\pi[-(\eta_2-ix)^2+
r^2]^3}\right].
\end{eqnarray}
To treat these poles, we should proceed as in the case of the first
integral, 
as indicated in Fig. (\ref{figos3}). Now the poles are located at 
$\eta_2= \pm r +ix$ (third order), and at $\eta_2 = \pm ix$ (first order).
Thus, we obtain the result,

\begin{eqnarray}\label{eq13}
I_2 = \frac{-4}{ r^4[2a_0^2x - a_1^2\omega\sinh(\omega x)] } +
\frac{2i}{2a_0^2x - a_1^2\omega\sinh(\omega x)} \Theta,
\end{eqnarray}
where $\Theta$ is given by:

\begin{figure}
\begin{center}
\includegraphics[width=0.50\textwidth]{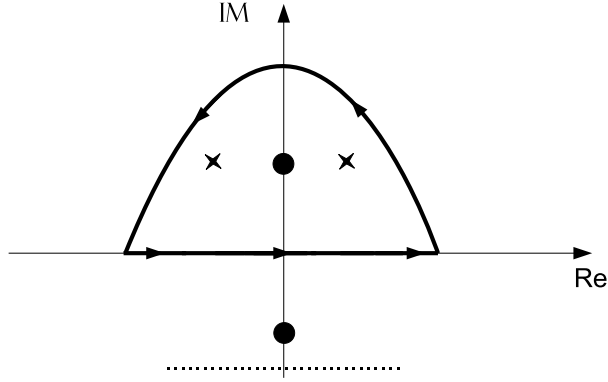}
\end{center}
\caption{This figure shows the easiest possible contour we can choose
  to 
integrate Eq. (\ref{eqos2}). Here the symbol $\bullet$ represents the 
first order poles located at $ix$, the $\textbf{X}$ represents the
third 
order poles, and $\cdot\cdot\cdot\cdot$ the
infinity first order poles at $\eta_{1k} = 
-\frac{i}{\omega}\ln\left(\alpha^2 B(\eta_{1k}) +
 \sqrt{\alpha^2 B(\eta_{1k}) -1 })\right) + \frac{1}{2}(2k +
1)\frac{\pi}{\omega}$. }\label{figos3}
\end{figure}

\begin{eqnarray}
 \Theta \equiv &&\frac{-\{2a_0^2(- r'+ix) -
a_1^2\omega\sinh[\omega(- r'+ix)]\}^2}{ r\{a_0^2(- r+ix)^2 +
a_0^2\eta_0^2 + a_1^2\cos[\omega(- r+ix)]\}^3}  \\
\nonumber && + \frac{2a_0^2 - a_1^2\omega^2\cos[\omega(-
r'+ix)]}{2r\{a_0^2(- r+ix)^2 + a_0^2\eta_0^2 + a_1^2\cos[\omega(-
r + ix)]\}^2 }
\\ \nonumber && + \frac{\{2a_0^2( r'+ix) -
a_1^2\omega\sinh[\omega( r'+ix)]\}^2}{ r\{a_0^2(
r+ix)^2 + a_0^2\eta_0^2 + a_1^2\cos[\omega( r+ix)]\}^3} \\
\nonumber &&+\frac{-\{2a_0^2 - a_1^2\omega^2\cos[\omega(
r'+ix)]\}}{2r\{a_0^2( r+ix)^2 + a_0^2\eta_0^2 + a_1^2\cos[\omega(
r + ix)]\}^2 }.
\end{eqnarray}
In order to check if our answer is correct, let us take 
$a_1 \simeq 0$ and $x = \eta_0$. Then we obtain,

\begin{equation}
I_2 = \frac{4\eta_0^2 - r'^2}{\eta_0^2 a_0^4[4\eta_0^2 +  r^2]^3},
\end{equation}
which is the same answer we found in the bouncing case Eq. (\ref{eqdeltau10}).
Now, using the smallest power terms in $a_1$ and 
$ r \ll \eta_0$, we have the squared coordinate velocity written as,

\begin{equation}
\langle\Delta u(r_1) \Delta u(r_2)\rangle \cong
\frac{q^2}{m^2}\left[\frac{3}{16a_0^4\eta_0^6} +
\frac{a_1^2}{2a_0^6 r^4\eta_0^3}\left(3\omega\sinh(\omega\eta_0) +
\frac{\cosh(\omega\eta_0)}{\eta_0} \right) \right],
\end{equation}
and when $n_f \gg n_0$, $\langle\Delta v(r_1)\Delta v(r_2)\rangle$ is given by Eq.~(\ref{eqosc}).

\end{document}